\documentclass{article}
\usepackage{spconf,amsmath,graphicx}

\title{Towards Improving Speaker Distance Estimation through Generative Impulse Response Augmentation}
\name{Anton Ratnarajah$^{1}$, Mehmet Ergezer$^{1,2}$, Arun Nair$^{1}$, Mrudula Athi$^{1}$}
\address{Amazon$^{1}$, Wentworth Institute of Technology$^{2}$}

\begin{document}
%
\maketitle
\begin{abstract}
The Room Acoustics and Speaker Distance Estimation (SDE) Challenge at ICASSP 2025 explores the effectiveness of augmented room impulse response (RIR) data for improving SDE model performance. This challenge at GenDARA involves generating RIRs to supplement sparse datasets and fine-tuning SDE models with the augmented data. We employ the open-source fast diffuse room impulse response generator (FastRIR) conditioned only on speaker and listener locations. We design a quality filter to ensure generated RIR alignment with challenge RIRs, and hyperparameter optimization is employed for model fine-tuning. Our approach reduces the mean absolute error (MAE) of the five positions from 1.66m to 0.6m for GWA rooms and from 2.18m to 0.69m for Treble rooms, with results demonstrating that the augmentation approach significantly improves estimation accuracy, particularly at medium to long distances.
\end{abstract}
\section{Introduction}
\label{sec:intro}
The Room Acoustics and Speaker Distance Estimation (SDE) Challenge at ICASSP 2025 aims to investigate the impact of augmented room impulse response (RIR) data on SDE model performance \cite{GenDA2025_RoomAcoustics}. This challenge at GenDARA involves two tasks: augmenting RIR data using generation systems and improving SDE models with the augmented data.

Speaker distance estimation is critical for various applications including smart speakers, teleconferencing systems, hearing aids, and spatial audio rendering. Accurate distance estimation enhances voice capture quality, improves speech recognition performance, and enables more realistic sound reproduction.

For Task 1, we implement a multi-stage training approach for our modified FAST-RIR model. Initially, we utilize 100,000 RIRs from the GWA dataset~\cite{tang2022gwa} for base training, with the modified FAST-RIR specifically conditioned only on speaker and listener positions. Given the distinct characteristics of Treble and GWA simulation methods, we conduct separate fine-tuning processes for each dataset. For each fine-tuning phase, we allocate 80\% of the respective enrollment RIRs for training, while reserving the remaining 20\% for model validation and optimal checkpoint selection in distance estimation tasks. This systematic approach ensures dataset-specific optimization while maintaining robust validation protocols.

Task 2 fine-tunes state-of-the-art SDE models \cite{neri2024speaker} using the generated RIRs, evaluating their effectiveness. Baseline experiments use the C4DM dataset \cite{stewart2010database} and VCTK dataset \cite{yamagishi2019cstr}. Participation in both tasks was encouraged to assess the generated RIRs' impact on SDE performance.

\section{Methodology}
\label{sec:methods}
In this section, we detail the methodology employed to enhance the performance of speaker distance estimation models using augmented room impulse response (RIR) data. Our approach consists of two primary tasks: augmenting RIR data with a RIR generation system and improving the speaker distance estimation model using the augmented data.

\subsection{Task 1: Augmenting RIR Data with RIR Generation System}
To generate a diverse set of Room Impulse Responses (RIRs), we employ the open-source Fast Diffuse Room Impulse Response Generator (FastRIR) ~\cite{Fast-RIR}. This tool enables the creation of synthetic RIRs that effectively simulate real-world acoustic environments. To meet our challenge's specific requirements, we modify FastRIR's conditional GAN architecture to condition solely on speaker and listener locations, ensuring the generated RIRs accurately represent varying speaker-listener distances.

Our modification to the FastRIR architecture includes:
\begin{itemize}
    \item Removal of room geometry conditioning to focus exclusively on source-receiver positioning.
    \item Extending the FAST-RIR Generator to Generate 1-Second Room Impulse Responses at 32 kHz Sampling Rate.
    \item Adaptation of the input feature representation to encode only distance-related parameters.
    \item Adpatation of RIR representation scheme proposed in MESH2IR~\cite{ratnarajah2022mesh2ir,listen2scene} to ensure consistent energy distribution across different distances.
\end{itemize}

Our training strategy follows a two-stage approach. First, we pre-train our network using 100,000 RIRs from diverse 3D environments in the GWA dataset ~\cite{tang2022gwa}. Subsequently, we implement separate fine-tuning processes for Treble and GWA enrollment data, utilizing 80\% of each dataset for training and the remaining 20\% for checkpoint selection. Given the limited enrollment RIRs, we make the practical assumption that acoustic conditions remain consistent within Rooms 1-10 and Rooms 11-20, with source and listener locations being the primary varying parameters. We maintain separate generative models for Treble and GWA enrollment RIRs due to their distinct simulation methodologies, as combining these disparate RIRs during fine-tuning with limited data could compromise model performance.

The curated high-quality RIRs were then used to fine-tune the state-of-the-art speaker distance estimation model (SDE) ~\cite{neri2024speaker}. This final fine-tuning step with the augmented dataset aims to enhance the model's robustness and generalization capabilities across diverse acoustic environments with varying speaker-listener distances.

\subsection{Task 2: Improving Speaker Distance Estimation Model with Augmented RIR Data}
Using our generative RIR model, we produced approximately 1 million RIRs by sampling various positions of speakers and listeners. However, the limited data for fine-tuning led some RIRs to show implausible reverberation times, mainly because of noise in their tails. To address this, a quality control process was executed to ensure consistency with the Treble RIRs' distribution.

Our quality filter implementation employed the following criteria:
\begin{itemize}
    \item T60 reverberation time within ±20\% of the reference distribution, excluding data with T60s exeeding 1.8695
    \item Direct-to-reverberant ratio (DRR) consistent with physical expectations at given distances, excluding data with source-receiver distances $<0.8$m and $>7.1$m.
    \item Energy decay curve shape matching typical acoustic behavior for the target environments
    \item Early reflection patterns consistent with real-world measurements
\end{itemize}

\begin{figure}[t]
\centering
\includegraphics[width=\linewidth]{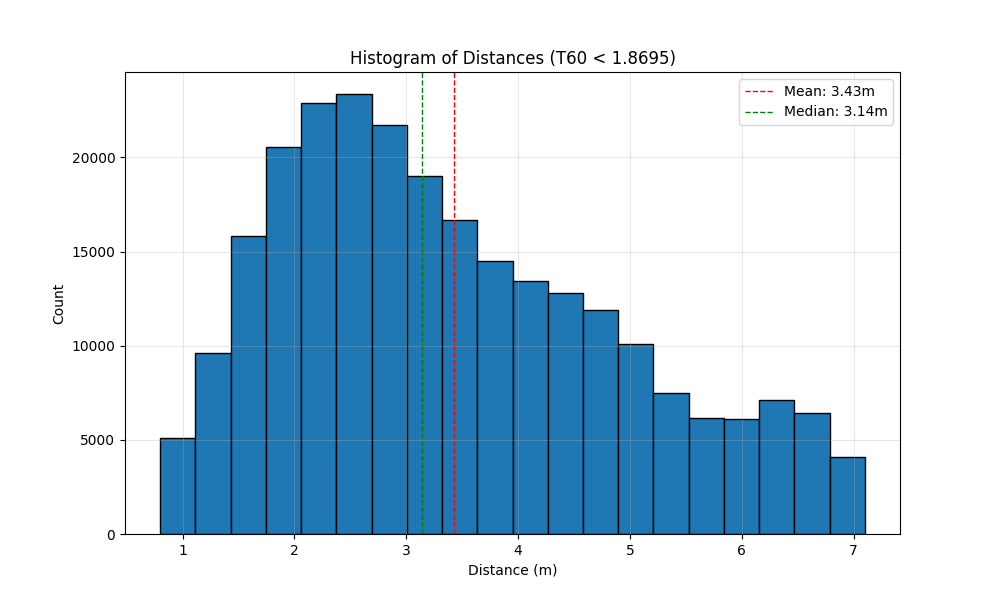}
\caption{Distribution of the speaker distances (in meters) for the RIRs used to fine-tune the SDE model. The distribution shows a concentration of samples between 1-5m, matching the expected range in typical indoor environments.}
\label{fig_filter}
\end{figure}

This filtering process resulted in a yield of approximately 25\%, equivalent to around 260,000 high-quality RIRs, retaining those with reverberation times and speaker distances that matched the distribution observed in Treble and GWA RIRs. Figure~\ref{fig_filter} illustrates the distance distribution for these refined RIRs. This selective method was crucial in preserving the quality of the dataset. These filtered RIRs were used for further tuning of the SDE.

To optimize the performance of the distance estimation model, we employ an off-the-shelf hyperparameter optimization algorithm. This helps us in systematically exploring the hyperparameter space to identify the optimal learning rate (ranging from 1e-5 to 1e-3), and epoch count (between 5 and 50). The goal is to minimize the test error on the provided Treble IRs while fine-tuning the model using our generated IRs.

Using the augmented data set and optimizing the hyperparameters, we achieve a more accurate and reliable speaker distance estimation model. This model is expected to perform well in various real-world scenarios, improving the overall performance of acoustic-based applications.

\section{Results}
\label{sec:results}
\begin{figure}[t]
\centering
\includegraphics[width=\linewidth]{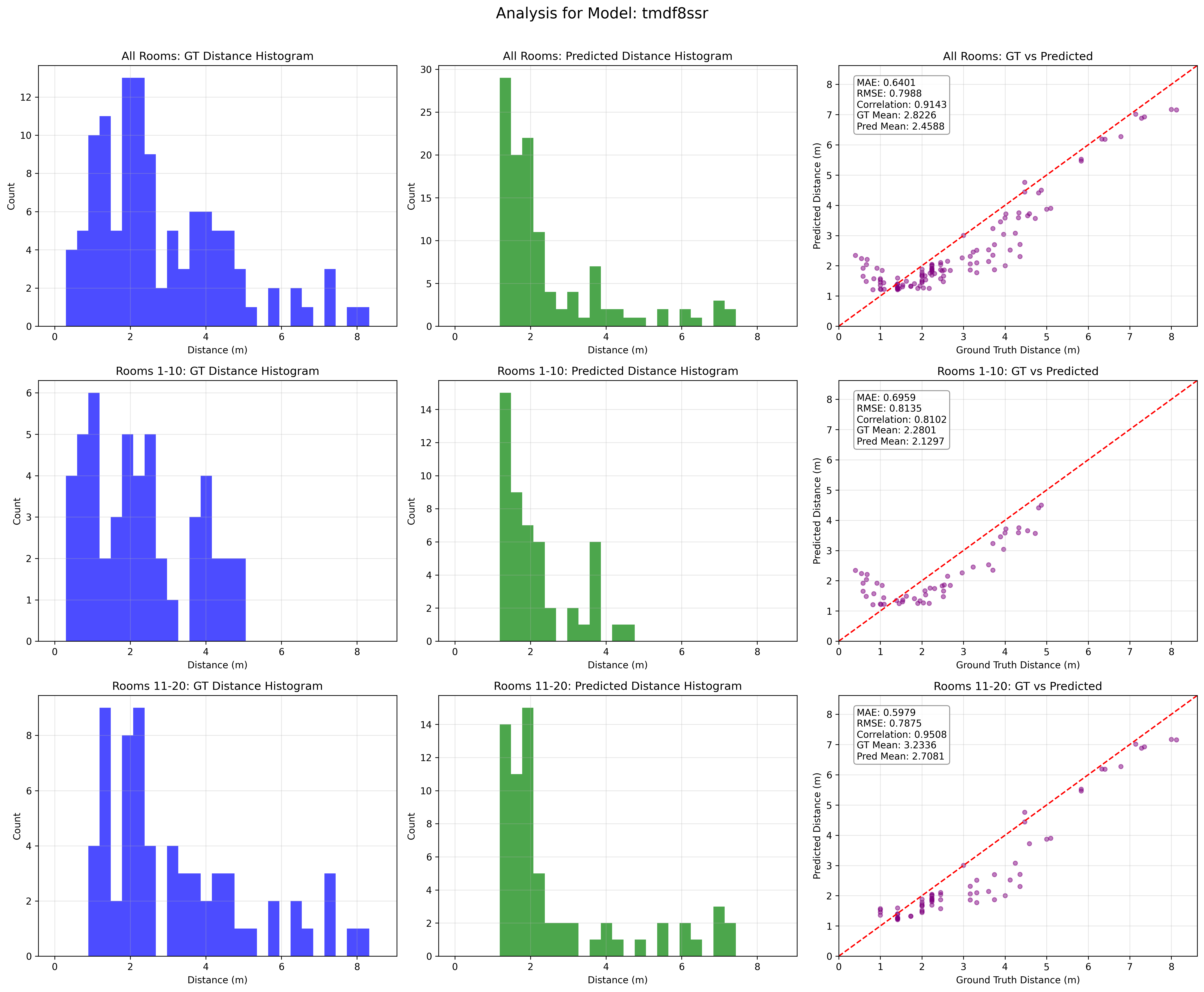}
\caption{Performance evaluation of our speaker distance estimation model. Left column: Ground truth distance distributions (in meters) for all test scenarios. Middle column: Distribution of predicted distances using our fine-tuned SDE model. Right column: Scatter plots comparing predicted vs. true distances with corresponding MAE values and correlation coefficients. The rows represent all rooms (top), Treble rooms 1-10 (middle), and GWA rooms 11-20 (bottom).}
\label{fig_results}
\end{figure}

Figure~\ref{fig_results} presents a comprehensive analysis of our SDE model's performance across all twenty test rooms, with separate evaluations for the first ten Treble rooms and the last ten rooms from GWA dataset. The figure is organized in three rows (all rooms, Treble rooms 1-10, and GWA rooms 11-20) and three columns (ground truth distance distributions, predicted distance distributions, and prediction accuracy scatter plots).

Analysis of the distance distributions in Figure~\ref{fig_results} reveals significant differences between the Treble and GWA datasets. Treble rooms predominantly feature speaker placements within a 5-meter range, with some instances positioned extremely close (under 1 meter). In contrast, GWA rooms exhibit a wider distribution extending up to 8 meters, and representation below the 1-meter threshold. This distinction in data distribution provides valuable insight into the characteristics of each dataset and offers context for interpreting model performance.

The prediction accuracy assessment in the third column demonstrates that our model achieves superior performance on GWA rooms, with a mean absolute error (MAE) of 0.6 meters compared to 0.69 meters for Treble rooms. The correlation coefficients further confirm this trend, with GWA predictions showing stronger alignment with ground truth values. When comparing our results to the beasline SDE model provided, this is still a significant improvement from the baseline of 1.66m and 2.18m on GWA and Treble rooms, respectively.

Error analysis reveals that distance prediction accuracy decreases significantly for speaker positions closer than 1 meter, particularly evident in the Treble room evaluations. This limitation stems from two factors: (1) our generative RIR model was trained on limited examples within this close-proximity range, and (2) acoustic phenomena at very close distances exhibit unique characteristics that are challenging to model without specialized training data. At distances beyond 1 meter, our model maintains consistent performance with errors typically around 0.5 meters.

\begin{figure}[t]
\centering
\includegraphics[width=1.5in]{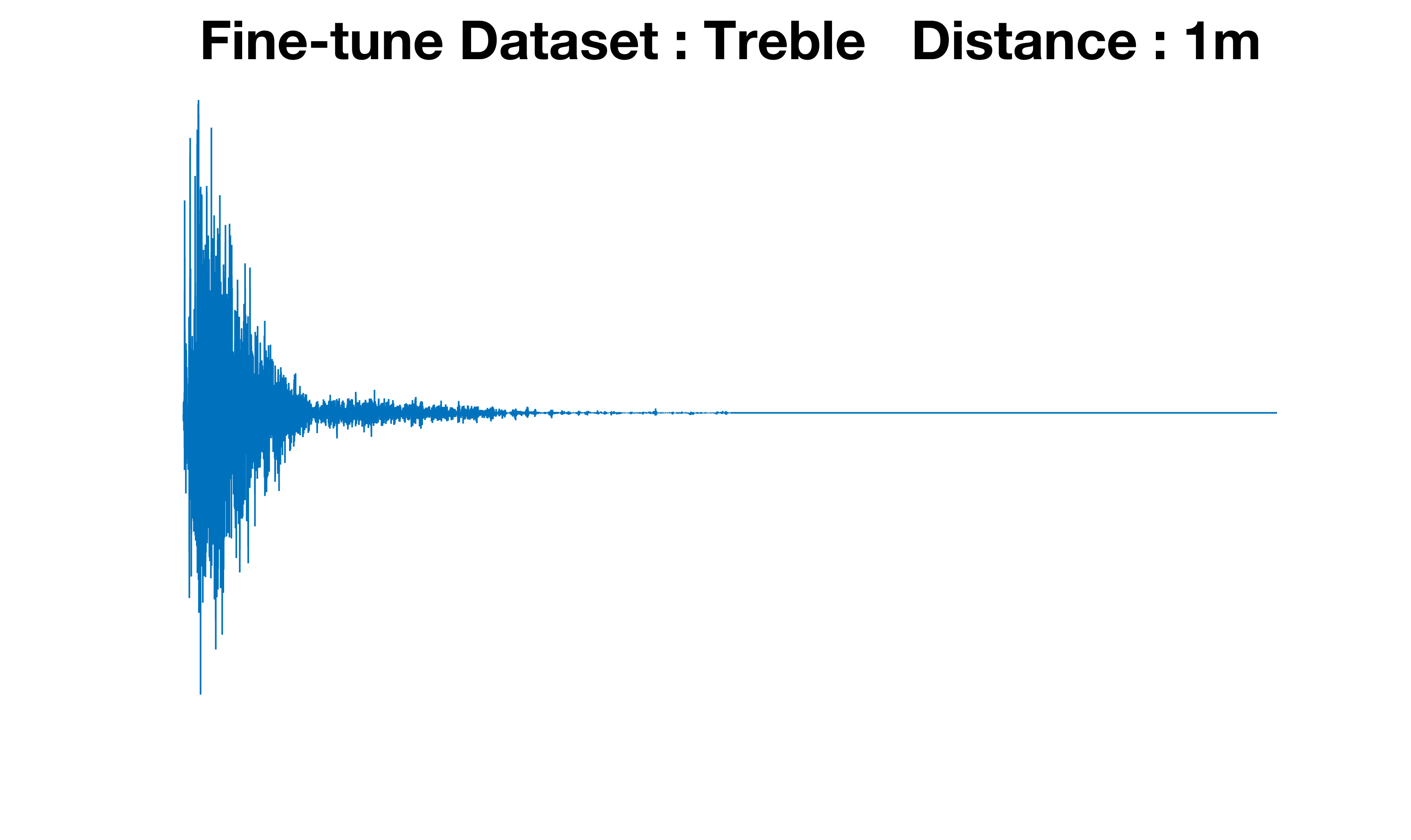}
\includegraphics[width=1.5in]{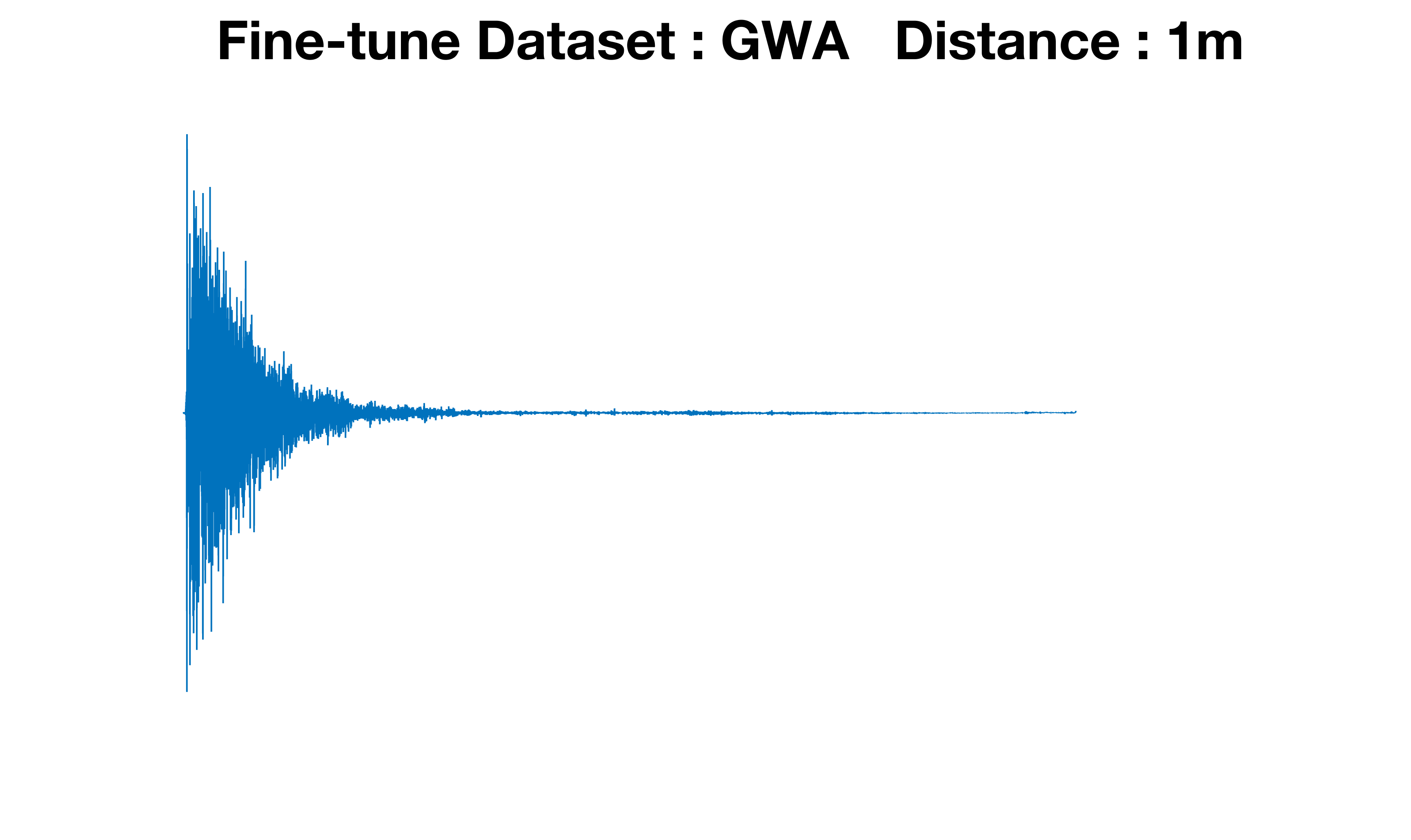}
\caption{Time-domain comparison of RIRs generated using modified FAST-RIR for test scenarios. Left: RIR generated using fine-tuned modified FAST-RIR with Treble enrollment data. Right: RIR generated using fine-tuned modified FAST-RIR with GWA enrollment data. The comparison reveals distinct reflection patterns between the modified FAST-RIR fine-tuned using two different datasets at identical source-receiver distances.}
\label{fig_3}
\end{figure}

We conducted additional experiments with dataset-specific models, training separate SDE systems exclusively on either Treble or GWA augmented data. These specialized models demonstrated further performance improvements, with MAE reductions of 10\% for Treble rooms and 5\% for GWA rooms compared to our unified model. This suggests that simulation-specific fine-tuning can yield additional benefits when deployment contexts are known in advance. Figure~\ref{fig_3} illustrates distinct reflection patterns in the RIRs generated by the modified FAST-RIR model, which was fine-tuned separately on the Treble and GWA datasets. This observation underscores the importance of simulation-specific fine-tuning, demonstrating its benefits for both RIR generation and speaker distance estimation tasks. 


\section{Conclusions}
The methodology employed in the Room Acoustics and Speaker Distance Estimation (SDE) Challenge at ICASSP 2025 demonstrates significant improvements in SDE model performance. By augmenting room impulse response (RIR) data using our modified FastRIR tool \cite{Fast-RIR} and fine-tuning the state-of-the-art distance model \cite{neri2024speaker} with the generated data, we achieved notable reductions in estimation error, with MAE values reducing from baselines of  1.66m to 0.6m and 2.18m to 0.69m for GWA and Treble rooms, respectively. 

Our rigorous quality control filtering process, retaining only 25\% of generated RIRs, proved critical in ensuring the realism and relevance of the augmented dataset. The hyperparameter optimization further enhanced model performance by identifying optimal learning rates and training durations specific to our augmented data characteristics.

This work makes several key contributions to the field: (1) demonstrating the effectiveness of conditioned generative models for RIR augmentation, (2) establishing a quality filtering methodology for synthetic acoustic data, and (3) providing empirical evidence that well-designed data augmentation significantly enhances distance estimation accuracy, particularly at medium to long distances.

Future research directions could explore combining multiple generative approaches, developing specialized models for extreme close-range distance estimation ($<1m$), and extending these techniques to more diverse acoustic environments beyond the controlled room settings evaluated in this challenge.

\vfill\pagebreak
\bibliographystyle{IEEEbib}
\bibliography{main2}

@INPROCEEDINGS{listen2scene,
  author={Ratnarajah, Anton and Manocha, Dinesh},
  booktitle={2024 IEEE Conference Virtual Reality and 3D User Interfaces (VR)}, 
  title={Listen2Scene: Interactive material-aware binaural sound propagation for reconstructed 3D scenes}, 
  year={2024},
  volume={},
  number={},
  pages={254-264},
  doi={10.1109/VR58804.2024.00048}}

@inproceedings{GenDA2025_RoomAcoustics,
  title={Generative Data Augmentation Challenge: Synthesis of Room Acoustics for Speaker Distance Estimation},
  author={Jackie Lin and Georg Gotz and Hermes Sampedro Llopis and Haukur Hafsteinsson and Steinar Guonsson and Daniel Gert Nielsen and Finnur Pind and Paris Smaragdis and Dinesh Manocha and John Hershey and Trausti Kristjansson and Minje Kim},
  booktitle={IEEE International Conference on Acoustics, Speech and Signal Processing Workshops(ICASSPW)},
  year={2025}
}

@INPROCEEDINGS{Fast-RIR, 
author={Ratnarajah, Anton and Zhang, Shi-Xiong and Yu, Meng and Tang, Zhenyu and Manocha, Dinesh and Yu, Dong}, 
booktitle={ICASSP 2022 - 2022 IEEE International Conference on Acoustics, Speech and Signal Processing (ICASSP)},
title={Fast-Rir: Fast Neural Diffuse Room Impulse Response Generator},
year={2022}, 
volume={},
number={},
pages={571-575},
doi={10.1109/ICASSP43922.2022.9747846}}

@article{ratnarajah2022mesh2ir,
  title={MESH2IR: Neural Acoustic Impulse Response Generator for Complex 3D Scenes},
  author={Ratnarajah, Anton and Tang, Zhenyu and Aralikatti, Rohith Chandrashekar and Manocha, Dinesh},
  journal={arXiv preprint arXiv:2205.09248},
  year={2022}
}

@inproceedings{tang2022gwa,
  title={GWA: A large high-quality acoustic dataset for audio processing},
  author={Tang, Zhenyu and Aralikatti, Rohith and Ratnarajah, Anton Jeran and Manocha, Dinesh},
  booktitle={ACM SIGGRAPH 2022 Conference Proceedings},
  pages={1--9},
  year={2022}
}

@article{neri2024speaker,
  title={Speaker Distance Estimation in Enclosures from Single-Channel Audio},
  author={Neri, Michael and Politis, Archontis and Krause, Daniel and Carli, Marco and Virtanen, Tuomas},
  journal={IEEE/ACM Transactions on Audio, Speech, and Language Processing},
  year={2024},
  publisher={IEEE}
}

@inproceedings{stewart2010database,
  title={Database of omnidirectional and B-format room impulse responses},
  author={Stewart, Rebecca and Sandler, Mark},
  booktitle={2010 IEEE International Conference on Acoustics, Speech and Signal Processing},
  pages={165--168},
  year={2010},
  organization={IEEE}
}

@article{yamagishi2019cstr,
  title={CSTR VCTK Corpus: English multi-speaker corpus for CSTR voice cloning toolkit (version 0.92)},
  author={Yamagishi, Junichi and Veaux, Christophe and MacDonald, Kirsten and others},
  journal={University of Edinburgh. The Centre for Speech Technology Research (CSTR)},
  pages={271--350},
  year={2019}
}
\end{document}